\documentclass[twocolumn,showpacs,preprintnumbers,amsmath,amssymb,floatfix,prb]{revtex4}

\usepackage[colorlinks,hypertex]{hyperref}

\usepackage{times}
\usepackage{epsfig}

\newcommand{\be}[1]{\begin{equation}\label{#1}}
\newcommand{\ee}{\end{equation}}   
\newcommand{\eq}[1]{(\ref{#1})}

\begin{document}

\title{
	An $\mathcal{O}(N^2)$ Approximation for Hydrodynamic
	Interactions in Brownian Dynamics Simulations
}

\author{Tiham\'{e}r Geyer}
\email[Author to whom correspondence should be addressed. Electronic 
	   mail: ]{Tihamer.Geyer@bioinformatik.uni-saarland.de}
\author{Uwe Winter}
\affiliation{Zentrum f\"ur Bioinformatik, Universit\"at des 
Saarlandes, D--66123 Saarbr\"ucken, Germany}

\begin{abstract}
	In the Ermak-McCammon algorithm for Brownian Dynamics, the
	hydrodynamic interactions (HI) between $N$ spherical particles are
	described by a $3N \times 3N$ diffusion tensor. This tensor has to
	be factorized at each timestep with a runtime of
	$\mathcal{O}(N^3)$, making the calculation of the correlated
	random displacements the bottleneck for many-particle simulations.
	Here we present a faster algorithm for this step, which is based
	on a truncated expansion of the hydrodynamic multi-particle
	correlations as two-body contributions. The comparison to the
	exact algorithm and to the Chebyshev approximation of Fixman
	verifies that for bead-spring polymers this approximation yields
	about 95\% of the hydrodynamic correlations at an improved runtime
	scaling of $\mathcal{O}(N^2)$ and a reduced memory footprint. The
	approximation is independent of the actual form of the
	hydrodynamic tensor and can be applied to arbitrary particle
	configurations. This now allows to include HI into large
	many-particle Brownian dynamics simulations, where until now the
	runtime scaling of the correlated random motion was prohibitive.
\end{abstract}

%

\maketitle

\section{Introduction}

For the simulation of diffusional processes on the scales of polymers
or proteins, Brownian Dynamics has proven to be a reliable workhorse 
\cite{GAB02,GRO03,GOR04,SPA06,HER06,DLUG07}. 
This coarse grained method builds upon Einstein's microscopic
explanation of the random motion of colloidal particles \cite{EIN05}
which had been observed earlier by the biologist Robert Brown when
studying pollen grains. In Einstein's explanation the solvent
molecules are replaced by a heat bath that models the collisions
between the solvent molecules and the much larger Brownian particles
by correctly distributed random forces. Together with the assumption
that the motion of the large observable particles is overdamped, i.e.,
that their velocities relax very fast compared to the time steps used
in the simulation, their diffusive motion is reproduced \cite{DHO96}.
Replacing the many small solvent molecules by a continuum dramatically
reduces the computational costs compared to an all-atom molecular
dynamics simulation, but for the price that also all interactions
between the Brownian particles have to be adapted to include the
effects of the now continuous solvent. For electrostatic interactions,
e.g., the polarizability of the solvent molecules and the
redistribution of the included ions can be described by an effective
shielding via the Debye-H\"uckel theory. When the solvent molecules
are not considered explicitly anymore, also so called hydrodynamic
interactions (HI) have to be included, which describe how the relative
motion of the Brownian particles is coupled mechanically via the
displaced solvent.

Now there is a fundamental difference between the effect of the
hydrodynamic coupling onto the direct interactions between the
Brownian particles and the one onto their random motion. An external
force acting on one of the particles accelerates this particle, which
in turn leads to a displacement of the other particles, too, mediated
by the displaced solvent molecules. Effectively, with hydrodynamics
the effect of the external forces is increased and the dynamics is
accelerated compared to a setup that does not consider hydrodynamics.
In contrast, the random motion of the Brownian particles models the
effect of the thermal fluctuations of the solvent molecules, the
strength of which depends on the temperature of the system.
Consequently, for a correct description of the overall diffusion the
random kicks must have the same strength with and without
hydrodynamics, i.e., whether the correlation due to the displaced
solvent is considered or neglected. Consequently, when hydrodynamic
interactions are included in a BD simulation, the random displacements
still have (up to second order) the same (temperature dependent)
magnitudes, but are correlated \cite{DEU71}. In a simulation, they now
have to be determined from a factorization of the diffusion tensor of
the complete system, which is numerically demanding.

In a BD simulation of $N$ particles using the Ermak-McCammon algorithm
\cite{ERM78}, the hydrodynamically correlated random displacements are
determined via a Cholesky factorization of the $3N \times 3N$
diffusion tensor, which results in a runtime of $\mathcal{O}(N^3)$,
while all the two-body interactions can be calculated in
$\mathcal{O}(N^2)$ runtime. For a many particle simulation with HI,
one is therefore spending most of the time evaluating the correlated
random displacements. For practical applications, this limits the
number of particles to a few dozens, when HI is included, while
without HI the dynamics of some thousand particles could be simulated
in the same time. As the already approximated direct interactions are
most important for the dynamics of most biologically interesting
association and dissociation processes, the time consuming HI is often
neglected \cite{MCG06}. For applications to polymers or the dynamics
of DNA, however, it may be important to explicitly include HI in order
to reproduce the correct dynamic behavior \cite{AHL01,ZHO06,SZY07}.

Considering that HI should on the one hand not be completely
neglected, but on the other hand is only one out of a handful of
interactions, there is an urgent need for faster algorithms to compute
the hydrodynamic coupling in the random motion of many particle
systems. Fixman proposed to use a Chebyshev approximation, which
scales as $\mathcal{O}(N^{2.5})$, to factorize the diffusion
coefficient \cite{FIX86b}. A drastically simplified approach to HI is
the effective mobility model by Heyes et al., which changes the
diffusion coefficients of the Brownian particles according to the
local density \cite{HEY95}. Such a model can obviously not reproduce
the full correlation between the individual particles. Based on these
ideas, Banchio and Brady developed an algorithm for infinite
homogenous suspensions that scales as $\mathcal{O}(N^{1.25} \log N)$
for large $N$ \cite {BAN03}. Due to its mathematical complexity, this
algorithm only performs well for very large numbers of particles.
Consequently, when HI is to be considered in BD simulations, many
current studies either use the original Cholesky factorization for 
smaller systems or Fixman's Chebyshev approximation with its 
better runtime scaling for larger many particle scenarios 
\cite{HUA02}.

Here we present a conceptionally different approach to tackle the
computationally expensive correlation of random motion in
many-particle simulations. Whereas all previous improvements to the
Ermak-McCammon algorithm only considered the factorization of the
hydrodynamic tensor, we argue that in typical BD simulations all
interactions are approximative anyhow. Consequently, even with the
exact (and very time-consuming) HI, the system dynamics will not be
exactly the same as in an experiment. Thus one may try to approximate
the correlations of the random forces, too, and reduce their
functional complexity without further perturbing the dynamics in the
simulation. The central requirement for such an approximation is thus
that it may not introduce any systematic drift.
 
This publication is organized as follows. Before we present our
truncated expansion ansatz, we shortly review the standard
Ermak-McCammon algorithm for BD simulations and how HI is treated
there. We also give a short introduction to the Chebyshev expansion of
the HI as introduced by Fixman. In section \ref{sec:Comparisons} we
compare our ansatz to the well established methods of Ermak and
Fixman, respectively. The main tests are the dynamics of a bead-spring
dimer, which was used by Ermak and McCammon to verify their algorithm,
and the behavior of bead-spring polymers of various lengths. The
simulation results show that our ansatz reproduces about 95\% of the
correlations due to hydrodynamics at a runtime which scales quadratic
with the bead number. We close with a summary and an outlook pointing
out potential extensions and applications.

\section{Evaluating hydrodynamic interactions}

\subsection{The Ermak-McCammon algorithm and the Rotne-Prager tensor}

In the often used Ermak-McCammon algorithm for Brownian Dynamics
\cite{ERM78, DIC84} the total displacement $\Delta r_i$ of the $i$th
particle during a timestep $\Delta t$ due to the external forces $F_j$
acting on all the particles and the random displacement $R_i$ is given by
\be{eq:BDDisplacement}
	\Delta r_i(\Delta t) = \sum_j \frac{D_{ij} F_j}{k_BT}\Delta t + 
	\sum_j \frac{\partial D_{ij}}{\partial r_j} \Delta t+ R_i(\Delta t)
\ee
The $3N \times 3N$ diffusion tensor $\mathbf{D} = (D_{ij})$ describes
the hydrodynamic coupling between the $N$ particles with their three
translational degrees of freedom. For the external forces,
$\mathbf{D}$ is used directly, whereas the hydrodynamically correlated
random displacements $R_i(\Delta t)$ are characterized only indirectly
by the statistical moments of a vanishing average and a finite
covariance, which is described by the corresponding entries of
$\mathbf{D}$:
\be{eq:Average}
	\langle R_i(\Delta t) \rangle = 0, \qquad
	\langle R_i(\Delta t) R_j (\Delta t) \rangle = 2D_{ij} \Delta t \, .
\ee
The most common forms for the diffusion tensor are the Oseen tensor
\cite{KIR48} and the Rotne-Prager-Yamakawa tensor \cite{ROT69, YAM70}.
For these approximations the second term of equation
\eq{eq:BDDisplacement}, which describes how the particles are dragged
into regions of faster diffusion, vanishes.

For identical spheres of radius $a$, the $3N \times 3N$
Rotne-Prager-Yamakawa hydrodynamic tensor, which couples the
translational displacements of the beads, consists of the following $3
\times 3$ submatrices $\mathbf{D}_{ij}$, where $i$ and $j$ label two
particles:
\begin{eqnarray}
	\mathbf{D}_{ii} & = & \frac{k_BT}{6\pi\eta a} \mathbf{I} \\[0.2cm]
	\mathbf{D}_{ij} & = & \frac{k_BT}{8\pi\eta r_{ij}}\left[
		(\mathbf{I} + \hat{r}_{ij} \otimes \hat{r}_{ij}) +
		\frac{2a^2}{3r_{ij}^2}
			(\mathbf{I} - 3\hat{r}_{ij} \otimes \hat{r}_{ij})
		\right]  \\ 
	& & \quad i\neq j \mbox{ and } r_{ij} \geq 2a 
		\label{eq:RPYfern}\\[0.5cm]
	\mathbf{D}_{ij} & = & \frac{k_B T}{6\pi\eta a} \left[
		\left(1 - \frac{9}{32} \frac{r_{ij}}{a}\right) \mathbf{I}
		+ \frac{3}{32}\frac{r_{ij}}{a}
			\hat{r}_{ij} \otimes \hat{r}_{ij}\right] \\
	& & \quad i\neq j \mbox{ and } r_{ij} < 2a
			\label{eq:RPYnah}
\end{eqnarray}
This tensor is positive definite for all particle configurations.
Dickinson et al. \cite{DIC84} later showed how rotational coupling can
be included, too. A generalization of the Rotne-Prager-Yamakawa tensor
for spheres of different radii was introduced by Garcia de la Torre
\cite{GAR77b}.

To determine the random displacements from equation \eq{eq:Average},
one needs to find $\mathbf{B}$ so that $\mathbf{D} =
\mathbf{B}\mathbf{B}^T$. One possible solution, which was used by
Ermak and McCammon, comes from a Cholesky factorization, giving
$\mathbf{B}$ as an upper (lower) tridiagonal matrix. The random
displacements are then determined as $\vec{R} = \mathbf{B} \vec{X}$
from a $3N$ dimensional vector $\vec{X}$ of normal distributed random
numbers.

\subsection{Fixman's Chebyshev approximation}

To avoid the computationally expensive Cholesky factorization of the
hydrodynamic tensor used in the Ermak-McCammon algorithm, Fixman
suggested in 1986 to approximate the square root of the diffusion
tensor via Chebyshev polynomials \cite{FIX86b}. Fortunately, the
explicit calculation of this matrix is not necessary and the vector of
correlated displacements can be determined iteratively by a series of
matrix-vector multiplications up to the order $L$ of the expansion:
\begin{eqnarray}
	\vec{R} & \approx & \sum_{l=0}^L a_l \vec{x}_l \\ 
	\vec{x}_0 & = & \vec{X} \\
	\vec{x}_1 & = & [d_a \mathbf{D} + d_b \mathbf{I}] \cdot \vec{X} \\
	\vec{x}_{l+1} & = & 2[d_a \mathbf{D} + d_b \mathbf{I}]\cdot 
		\vec{x}_l - \vec{x}_{l-1}
		\label{eq:FixmanIteration}
\end{eqnarray}
The factors $d_a$ and $d_b$ are related to the range of eigenvectors 
$[\lambda_{min}, \lambda_{max}]$ of $\mathbf{D}$ as
\be{eq:dadbDef}
	d_a = \frac{2}{\lambda_{max} - \lambda_{min}} 
	\quad \mbox{ and } \quad
	d_b = \frac{\lambda_{max} + \lambda_{min}}{\lambda_{max} - \lambda_{min}}
\ee
For further details like the evaluation of the expansion coefficients
$a_l$ we refer the readers to, e.g., Press et al.\cite{PRE97}

To determine the necessary order $L$ of the Chebyshev approximation
and to control whether the eigenvalue spectrum of $\mathbf{D}$ is
bounded correctly, we followed the procedure of Jendrejack et al.
\cite{JEN00}. They introduced the relative error $\epsilon_f$ ($E_f$
in their notation) derived from the approximated random displacements
$\vec{R}$, the uncorrelated random numbers $\vec{X}$, and the
diffusion tensor $\mathbf{D}$:
\be{eq:FixmanEps}
	\epsilon_f = \sqrt{\frac{|
		\vec{R}\cdot \vec{R} - \vec{X}\mathbf{D}\vec{X}|}
		{\vec{X}\mathbf{D}\vec{X}}}
\ee
As $\vec{R}\cdot \vec{R}$ can be evaluated with negligible additional
cost, we used $\epsilon_f$ to monitor the convergence of the Chebyshev
iteration \eq{eq:FixmanIteration}. When $\epsilon_f$ remained above
the chosen threshold with the actual values of $L$, $\lambda_{min}$,
and $\lambda_{max}$, $L$ was increased by three and $\lambda_{min}$
and $\lambda_{max}$ were recalculated. For the next iteration, $L =
l_{max} + 3$ was used, where $l_{max}$ is the index of the last term
of the iteration \eq{eq:FixmanIteration} required to get $\epsilon_f$
below the chosen threshold. Following Jendrejack et al. \cite{JEN00},
$\epsilon_f$ was set to $10^{-3}$ if not otherwise noted. For this
value the numerical results were sufficiently close to the results
with the exact Cholesky factorization.

\subsection{The truncated expansion ansatz}

The two methods outlined above focus on factorizing $\mathbf{D}$.
Compared to the exact Cholesky factorization, the numerical
approximation of Fixman manages to efficiently get close to the exact
value of the square root of the diffusion tensor. We now present an
approximation tailored for practical applications of many particle
simulations, which algorithmically treats the displacements from the
external forces and from the random forces on an equal footing.

For such an approximation we start from equation
\eq{eq:BDDisplacement}. Neglecting the random displacements, the
displacement along coordinate $i$ during $\Delta t$ is
\be{eq:DRohneZufall}
	\Delta r_i(\Delta t) = 
		\sum_j \frac{D_{ij} F_j}{k_BT}\Delta t  = 
		\frac{D_{ii} \Delta t}{k_BT} F_i^{eff},
\ee
where we introduced the hydrodynamically corrected effective force
$F_i^{eff}$:
\be{eq:Feffective}
	F_i^{eff} =  \sum_j \frac{D_{ij}}{D_{ii}} F_j
\ee
This reformulation is independent of the actual form of the
hydrodynamic tensor.

For the random displacements, we now make an ansatz with the same
structure, i.e., we also introduce a hydrodynamically corrected random
force $f_i^{eff}$ acting on coordinate $i$ which is derived from the
uncorrelated random forces $f_j$ that would act on each of the
particles in the absence of HI. In this ansatz the displacements due
to the random forces alone are given as
\be{eq:Ansatz}
	\Delta r_i(\Delta t) = \frac{D_{ii} \Delta t}{k_BT} C_i
		\sum_j \beta_{ij} \frac{D_{ij}}{D_{ii}} f_j
		= \frac{D_{ii} \Delta t}{k_BT} f_i^{eff}.
\ee
The structure of this ansatz, which effectively factorizes
$\mathbf{D}$ into individual two-body contributions, is taken from
equation \eq{eq:DRohneZufall}. The scaling factor $C_i$ takes care
that for each individual coordinate its unperturbed diffusion
coefficient $D_{ii}$ is regained \cite{DEU71}, while $\beta_{ij}$
allows for different weights when $\sqrt{\mathbf{D}}$ is used instead
of $\mathbf{D}$ in equation \eq{eq:Feffective}. As we will see later,
there is an individual scaling factor $C_i$ for each of the
coordinates, while, due to symmetry reasons, for the coefficients
$\beta_{ij}$ we only need two different values for the diagonal
$\beta_{ii}$ and for the off-diagonal $\beta_{ij}$ with $i \neq j$,
respectively.

Now the parameters $C_i$ and $\beta_{ij}$ have to be determined such
that the moments of the correlated displacements \eq{eq:Average} are
reproduced with only small deviations. For the uncoupled random forces
we use
\be{eq:RandomAvg} 
	\langle f_i \rangle = 0 \quad \mbox{and}\quad
	\langle f_i\; f_j\rangle = \frac{2(k_BT)^2}{D_{ii} \Delta t} \delta_{ij} 
	\, ,
\ee
i.e., they reproduce the mean and covariance of the random
displacements in the uncorrelated case.

With correlation, i.e., with hydrodynamic coupling, the vanishing
average $\langle R_i\rangle$ is fulfilled straightforwardly from
$\langle f_i \rangle = 0$. For the covariance we start from the
product of $\Delta r_i$ and $\Delta r_j$ according to equation
\eq{eq:Ansatz}:
\be{eq:DRProdukt}
	\Delta r_i \Delta r_j = \left( \frac{\Delta t}{k_BT}\right)^2
		C_i C_j \sum_{k, l} \beta_{ik} \beta_{jl} D_{ik} D_{jl}
		f_k f_l
\ee
Inserting \eq{eq:RandomAvg} leads to the condition
\be{eq:DRCovariance}
	\langle r_i r_j \rangle = 
		2 \Delta t C_i C_j \sum_k \beta_{ik} \beta_{jk} 
		\frac{D_{ik} D_{jk}}{D_{kk}} 
	\stackrel{!}{=} 2 D_{ij} \Delta t .
\ee
The terms with $k \neq l$ drop from the double sum, because the $f_j$
are uncorrelated. Requiring that the variance of equation
\eq{eq:Average} be reproduced \cite{DEU71} for $i=j$ allows us to
determine the normalization constants $C_i$:
\be{eq:Normalisation}
	\left(\frac{1}{C_i}\right)^2 =
		\sum_k \beta_{ik}^2 \frac{D_{ik}^2}{D_{ii}D_{kk}}
\ee
Without loss of generality, we can set $\beta_{ii} = 1$. Then,
equation \eq{eq:Ansatz} reduces to the usual form in the limit of
vanishing HI. In this case, also $C_i = 1$ and the displacement
$\Delta r_i(\Delta t)$ can be calculated with equation
\eq{eq:DRohneZufall} from the sum of the random force $f_i$ and the
external force $F_i$.

With $\beta_{ii} = 1$, equation \eq{eq:Normalisation} can also be
written as
\be{eq:Normalisation2}
	\left(\frac{1}{C_i}\right)^2 =
		1 + \sum_{k \neq i} \beta_{ik}^2 \frac{D_{ik}^2}{D_{ii}D_{kk}}\, ,
\ee
where the term with $k=i$ was taken out of the sum.

With our ansatz \eq{eq:Ansatz} there is no set of coefficients which
can be determined numerically efficiently, with which equations
\eq{eq:Average} can be fulfilled simultaneously. To determine the
remaining off-diagonal coefficients $\beta_{ij}$, we therefore proceed
by assuming that the hydrodynamic coupling is weak, i.e., that the
off-diagonal entries $D_{ij}$ of the hydrodynamic tensor are much
smaller than the individual diffusion coefficients $D_{ii}$. Then we
can use $\sqrt{1+\epsilon} \approx 1 + \epsilon/2$ to expand equation
\eq{eq:Normalisation2}:
\be{eq:EtwaNormalisierung}
	\frac{1}{C_i} \approx 1 + \frac{1}{2}
		\sum_{k \neq i} \beta_{ik}^2 \frac{D_{ik}^2}{D_{ii}D_{kk}}
\ee
Thus, the product $(C_i C_j)^{-1}$ in equation \ref{eq:DRCovariance}
can be approximated as
\be{eq:EtwaVorfaktor}
	\frac{1}{C_i C_j} \approx 1 + 
	\frac{1}{2} \sum_{k\neq i} 
	\frac{\beta_{ik}^2 D_{ik}^2}{D_{ii}D_{kk}} +
	\frac{1}{2} \sum_{l\neq j} \frac{\beta_{jl}^2 D_{jl}^2}{D_{jj}D_{ll}}
	+ \mathcal{O}\left( \frac{D_{ij}}{D_{ii}} \right)^4
\ee
i.e., a constant term plus terms that are quadratic and quartic in
$D_{ij}/D_{ii}$. As we have the same $\beta_{ii} = \beta = 1$ for all
$i$, we also set all $\beta_{ij}$ for $i \neq j$ to the same value
$\beta'$. Without the quartic term, equation \eq{eq:EtwaVorfaktor}
then becomes
\be{eq:VorfaktorKurz}
	\frac{1}{C_i C_j} \approx 1 + (N-1)\,\beta'^2 \epsilon^2 
\ee
with the averaged relative coupling strength $\epsilon = \langle
D_{ij} / D_{ii}\rangle$.

Starting directly from equation \eq{eq:DRCovariance} we get the 
non-approximated relation
\be{eq:Bedingung}
	\frac{1}{C_i} \frac{1}{C_j} = \sum_k
		\beta_{ik} \beta_{jk}\frac{D_{ik}D_{jk}}{D_{kk}D_{ij}}.
\ee
On the rhs of this equation the two terms with $k=i$ and $k=j$ can be
taken out of the sum. They are independent of the hydrodynamic
coupling, while all other terms are of first order in $D_{ij}
/D_{ii}$. With $\beta = 1$ this yields:
\be{eq:Bedingung2}
	\frac{1}{C_i} \frac{1}{C_j} = 2 \beta' + (N-2) \beta'^2 
	\epsilon
\ee
Comparing equations \eq{eq:VorfaktorKurz} and \eq{eq:Bedingung2} then
gives the quadratic equation
\be{eq:BetaQuadratisch}
	\beta'^2 \left[ (N-1) \epsilon^2 - (N-2) \epsilon \right] - 
	2\beta' + 1 = 0,
\ee
which can be solved for $\beta'$:
\be{eq:BetaIJ}
	\beta_{ij} = \beta' =  
		\frac{1 - \sqrt{1 - [ (N-1) \epsilon^2 - (N-2) \epsilon]}}
		{(N-1) \epsilon^2 - (N-2) \epsilon}
\ee
We note that $\beta_{ij}$ according to the above equation is not
defined in the case of absent HI. However, it converges to $\beta_{ij} =
1/2$ for vanishing HI, i.e., for $\epsilon \to 0$, the value obtained
from equation \eq{eq:BetaQuadratisch} for $\epsilon = 0$.

The resulting approximation to the hydrodynamically coupled random
displacements of equation \eq{eq:Ansatz} with the normalization
constants $C_i$ given by \eq{eq:Normalisation} and the two weights
$\beta_{ii} = 1$ and $\beta_{ij}$ according to equation \eq{eq:BetaIJ}
has the same structure as the displacements due to the
hydrodynamically coupled external forces in equation
\eq{eq:DRohneZufall}, resulting in the same runtime scaling of
$\mathcal{O}(N^2)$ for both the deterministic and the random
contributions to $\Delta r_i$. Consequently, with this algorithm,
which treats the external and the random forces on equal footing, HI
can now be included in all those BD simulations where the direct
interparticle forces can be computed---given our approximation is
accurate enough for the chosen application.

\subsection{Simulation details}

To evaluate how our ansatz compares to the standard methods, we ran BD
simulations of bead-spring polymers with $N=2 \ldots 2000$ beads of
radius $a$. The diffusion coefficient of the individual beads was set
to $D_0 = 1$, thus defining the time scale. The beads were connected
to their direct neighbors in the chain by harmonic springs with
\be{eq:Federn}
	V_h(x) = a_h (x - L)^2 .
\ee
The potential minimum was varied for the dimer simulations and set to
$L=3a$ for the polymer simulations. To prevent the beads from
overlapping, a repulsive harmonic potential between all beads was used
analogous to the setup of Ermak and McCammon \cite{ERM78}:
\be{eq:Softcore}
	V_c(x) = a_h(x - 2a)^2 \quad \mbox{ for } \quad r < 2a
\ee
The spring constant for both interactions was set to the rather stiff
value of $a_h = 50$ $k_BT/a^2$, with which we used a conservatively
estimated integration timestep $\Delta t = 0.001$.

The Cholesky factorization, the eigenvalue calculation, the matrix and
vector operations as well as the generation of the random numbers were
performed with the respective subroutines from the GNU scientific
library \cite{GSL06}.

For simulations of shorter polymers of $N < 70$, all simulation were
started with the beads of the polymer aligned along the $x$ axis with
a mutual separation $L=3$. Data analysis was started when the polymer
had reached its coiled state. This transition was observed via the
radius of gyration. For polymers with $N > 70$, the equilibration
was too slow with hydrodynamics included. For every chain length we
therefore started a set of simulations without hydrodynamics and a
longer timestep of $\Delta t = 0.003$ and saved the final positions
when the polymer had coiled up. These snapshots were then used as
starting points for simulations with the different forms of HI.

\section{Comparison of the methods}
\label{sec:Comparisons}

In the following section we evaluate how our truncated expansion
approximation to HI (TEA-HI) compares to the other methods, namely the
exact factorization of Ermak and McCammon and the Chebyshev
approximation introduced by Fixman. For this, we analyzed the
correlation coefficient of two random displacements in one dimension,
the dynamics of a dimer, which had been introduced as a test case by
Ermak and McCammon \cite{ERM78}, and several static and dynamic
properties of bead-spring polymers. These comparisons will confirm
that most of the hydrodynamic interaction is obtained with our
approximation at a greatly improved runtime scaling.

\subsection{Correlation coefficient}

The simplest test is to directly evaluate the correlation coefficient
$\rho_{ij}$ of two one-dimensional displacements that are
hydrodynamically coupled:
\be{eq:CorrCoeff}
	\rho_{ij} = \frac{\langle \Delta r_i \, \Delta r_j \rangle}
		{\langle \Delta r_i \rangle \langle \Delta r_j \rangle}
\ee

For $D_{11} = D_{22}$, $D_{12} = D_{21}$ was varied between 0 and
$D_{11}$ and $\rho_{ij}$ was averaged for correlated displacements
using both the explicit Cholesky factorization and our approximation.
For the whole range of $D_{12}$, our TEA-HI gave values for
$\rho_{12}$, which were by less than 0.1\% smaller than with the exact
factorization. Similar minor deviations were obtained, too, for all
tested cases with $D_{11} \neq D_{22}$.

\subsection{Dimer dynamics}

\begin{table}[t]
	\centering
	\begin{tabular}{cccccccccc}
		\hline \hline
		$L$ & $\langle d\rangle$ & &  & $D_{CM}$ & & & & $\alpha_D$ & \\
		& & \hspace{0.5cm} &Geyer& \quad Ermak\quad & theo. & 
			\hspace{0.5cm} &Geyer & \quad Ermak \quad & theo \\
		\hline
		2 & 2.028 & & 0.7164 & 0.7468 & 0.7464 & & 2.43 & 2.28 & 2.28\\
		3 & 3.003 & & 0.6530 & 0.6665 & 0.6665 & & 2.97 & 2.93 & 2.93\\
		4 & 4.002 & & 0.6175 & 0.6253 & 0.6249 & & 3.27 & 3.27 & 3.22\\
		8 & 8.001 & & 0.5612 & 0.5627 & 0.5625 & & 3.59 & 3.59 & 3.62\\
		20 & 20   & & 0.5247 & 0.5243 & 0.5250 & & 3.85 & 3.92 & 3.85\\
		66.7 &66.7& & 0.5073 & 0.5073 & 0.5075 & & 3.92 & 3.91 & 3.96\\
		\hline \hline
	\end{tabular}
	\caption{Comparison of the center-of-mass diffusion coefficient
	$D_{CM}$ and the rescaled correlation time $\alpha_D$ of the
	orientation of a dimer of spherical beads with radius $a$
	connected by a spring of length $L$ from BD simulations with HI to
	their respective theoretical values. The actual average separation
	during the simulation is given by $\langle d\rangle$ and was used
	to calculate the theoretical predictions from \eq{eq:DimerDCM} and
	\eq{eq:DimerAlpha}. The headings ``Geyer'' and ``Ermak'' denote
	the respective type of HI that was used in the simulations.}
	\label{tbl:Dimer}
\end{table}

In their original work \cite{ERM78}, Ermak and McCammon verified their
approach to BD with HI by comparing the numerically determined
diffusion coefficient of the center of mass of a dimer of spherical
beads with radius $a$ and separation $L$, $D_{CM}$, and the inverse of
the rescaled relaxation time of its orientation, $\alpha_D$, to
analytical results. Only translational coupling between the beads was
considered with the Rotne-Prager tensor. According to Ermak and
McCammon \cite{ERM78}, with this form of the HI, the analytical values
for $D_{CM}$ and $\alpha_D = T_D / \tau_D$ are
\begin{eqnarray}
	D_{CM} & = & \frac{D_0}{2} \left(1 + \frac{a}{d}\right) 
		\label{eq:DimerDCM}\\
	\alpha_D & = & 4 \left( 1 - \frac{3a}{4d} - 
		\frac{1}{2}\left(\frac{a}{d}\right)^3 \right)
		\label{eq:DimerAlpha}
\end{eqnarray}
Here, $d$ is the distance between the centers of the two beads.
$D_{CM}$ is given in units of the diffusion coefficient $D_0$ of a
single bead. The relaxation time $\tau_D$ is given in units of $T_D =
d^2/D_0$, which is the time that a single bead would need to diffuse
over the separation $d$ between the two beads of the dimer. Without
HI, the limiting values are $D_{CM} = 0.5$ and $\alpha_D = 4$.

In a spring-bead dimer the average distance between the two beads is
slightly larger than the length $L$ of the spring connecting them. We
therefore calculated $D_{CM}$ and $\alpha_D$ from the observed average
separation $\langle d \rangle$ during the simulation for different
spring lengths $L$. As seen in table \ref{tbl:Dimer}, our results
reproduce the analytical values for both $D_{CM}$ and $\alpha_D$ quite
well. The hydrodynamic coupling tends to be underestimated by less
than 7\% for touching spheres, where the correlation is strongest, and
much less for larger separations. Underestimating the HI has the
consequence that the diffusion coefficient $D_{CM}$ is slightly
smaller than the theoretical value and the relaxation time is slightly
shorter, resulting in a larger $\alpha_D$. For comparison, table
\ref{tbl:Dimer} also gives the numerical results for $D_{CM}$ and
$\alpha_D$ with the correct factorization of Ermak and McCammon. Their
deviation from the theoretical predictions is less than 2\% which
indicates the numerical uncertainties of the simulation results.

\subsection{Static properties of bead-spring polymers}

\begin{figure}[t]
	\begin{center}
		\includegraphics[]{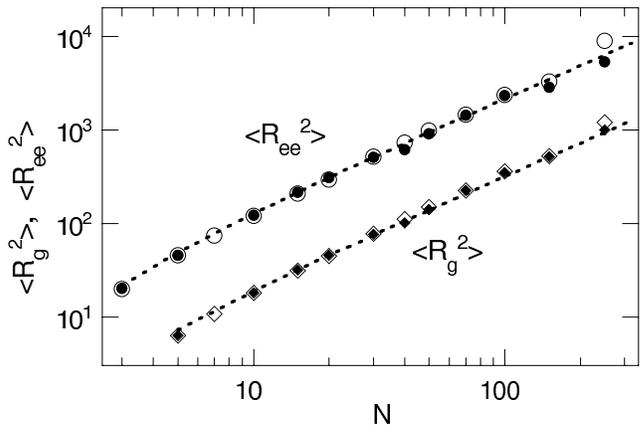}
	\end{center}
	\caption{Average radius of gyration $\langle R_g^2 \rangle$
	(diamonds) and end-to-end distance $\langle R_{ee}^2\rangle$
	(points) for polymers of various chain lengths $N$ from
	simulations where HI was calculated with Fixman's Chebyshev
	approximation (open symbols) and with our truncated expansion
	approximation (filled symbols), respectively. The lines indicate
	the theoretically predicted scaling $\propto (N-1)^{1.176}$.}
	\label{fig:RgRee}
\end{figure}

In the next two sections we present results from simulations of
bead-spring polymers of various chain length $N$. Here we look at the
equilibrium values of the radius of gyration $\langle R_g^2\rangle$
and the end-to-end $\langle R_{ee}^2\rangle$ distance. These are given
by
\be{eq:RGyration}
	\langle R_g^2 \rangle = 
		\frac{1}{2N^2} \sum_{ij} \langle r_{ij}\rangle
\ee
and
\be{eq:Ree}
	\langle R_{ee}^2\rangle = \langle (\vec{r}_N - \vec{r}_1)^2 
	\rangle .
\ee
Their theoretically predicted scaling behavior is
\be{eq:RgReeScaling}
	\langle R_g^2 \rangle \propto \langle R_{ee}^2\rangle 
		\propto (N-1)^{2\nu}
\ee
In a good solvent, perturbation analysis \cite{LI95,DUN02} predicts an
exponent of $\nu \approx 0.588$ for $N \to \infty$, which is also
reproduced in our simulations as shown in figure \ref{fig:RgRee}. For
clarity, figure \ref{fig:RgRee} only reproduces the results obtained
with Fixman's Chebyshev approximation and with our TEA-HI. Using the
original Cholesky factorization of Ermak and McCammon or no HI at all
gave results that were, within the numerical fluctuations,
indistinguishable from the ones shown. However, for runtime reasons we
only ran simulations for $N\leq 70$ with Ermak's original method.

Over the range of chain lengths $N=4 \ldots 200$, we obtained a ratio
of $\langle R_{ee}^2\rangle / \langle R_g^2\rangle \approx 6.8$, which
is about 10\% larger than the results of Li et al. \cite{LI95},
Jendrejack et al. \cite{JEN00}, or Liu and D\"unweg \cite{LIU03}. The
difference may be due to the fact that in our simulations the ratio
between bond length $L$ and bead radius $a$ is smaller than used
there. Consequently, the polymer can not be compacted as much as with
relatively smaller beads.

Even as the correct $\langle R_g^2\rangle$ and $\langle
R_{ee}^2\rangle$ do not directly prove that our truncated expansion HI
is correct, they show that it does not introduce any static
perturbations to the polymer conformations.

\subsection{Dynamic measures of bead-spring polymers}

\begin{figure}[t]
	\begin{center}
		\includegraphics[]{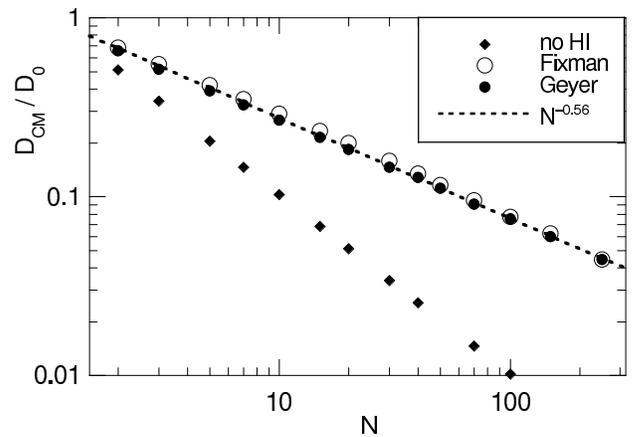}
	\end{center}
	\caption{Diffusion coefficient $D_{cm}$ of the center of mass of
	bead-spring polymers of various chain lengths $N$ from simulations
	without HI (diamonds) and with HI using Fixman's Chebyshev
	approximation (open points) or our truncated expansion
	approximation (filled points). The dashed line indicates the
	observed scaling of $N^{-0.56}$.}
	\label{fig:DCM}
\end{figure}

A central dynamic property of a diffusing polymer is its center of
mass diffusion coefficient $D_{cm}$, which is predicted to scale as
$D_{cm} \propto N^{-\nu}$ with $\nu \approx 0.588$ with Rotne-Prager
HI and $\propto 1/N$ without HI. The scaling without HI was reproduced
in our simulations, as can be seen in figure \ref{fig:DCM}, which
gives $D_{cm}$ in units of the diffusion coefficient $D_0$ of an
individual bead. Without HI, the results are only shown up to $N=100$,
but were calculated for $N=2\dots 400$.

With HI, the results with the original method of Ermak and McCammon
and with Fixman's approximation for $N=2\ldots 70$ were
indistinguishable within the numerical uncertainties. For $N > 70$,
only the faster Chebyshev approximation was used. $D_{cm}$ from these
simulation can be fitted well with $N^{-0.56}$, which means that in
our simulations the diffusion of long polymers was slightly faster
than predicted theoretically.

The results with our TEA-HI show the same scaling behavior, while
$D_{cm}$ is slightly smaller than with the correct factorization of
$\mathbf{D}$. The relative deviation between the results with our
approach vs. the Chebyshev approximation is about 5\%, i.e., our
truncated expansion reproduces about 95\% of the effect of the
hydrodynamic correlation on the overall diffusive motion of the
polymer.

\begin{figure}[t]
	\begin{center}
		\includegraphics[]{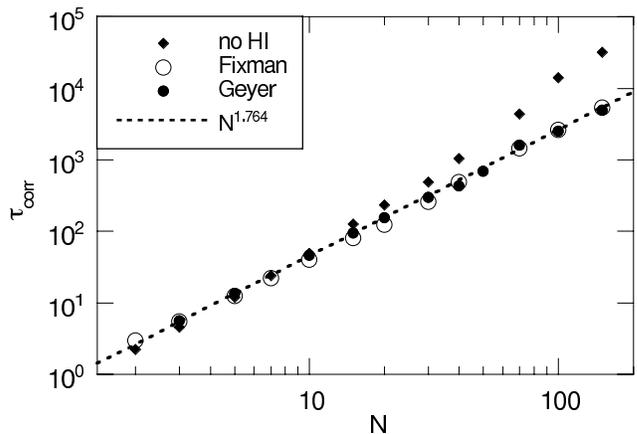}
	\end{center}
	\caption{Relaxation times $\tau_{corr}$ of the autocorrelation
	function of the end-to-end vector $\vec{R}_{ee}$ of polymers of
	various lengths $N$ from BD simulations without HI (diamonds) and
	with HI according to Fixman's Chebyshev approximation (open
	points) and our truncated expansion approximation (filled points),
	respectively. The relaxation times were obtained by fitting a
	stretched exponential to the autocorrelation function. The dashed
	line indicates the scaling $\propto N^{3 \nu}$ with $\nu =
	0.588$.}
	\label{fig:korrelations}
\end{figure}

Another measure, which is sensitive to the internal dynamics of the
polymer, is the autocorrelation function $\langle \vec{R}_{ee}(t)
\cdot \vec{R}_{ee}(0) \rangle$ of the end-to-end vector
$\vec{R}_{ee}$, which decays exponentially with a time constant
$\tau_{corr}$ (the Zimm time). Figure \ref{fig:korrelations} shows
that the fitted relaxation times are proportional to $N^{3\nu}$ as
predicted theoretically. With our TEA-HI, the relaxation is again
slightly slower by some 10\%, which indicates that our approximation
takes most of the hydrodynamic correlation between the beads into
account. Without HI, the relaxation of $\vec{R}_{ee}$ takes about one
order of magnitude longer at $N=100$ and even more for longer chains.

\subsection{Runtime considerations}

\begin{figure}[t]
	\begin{center}
		\includegraphics[]{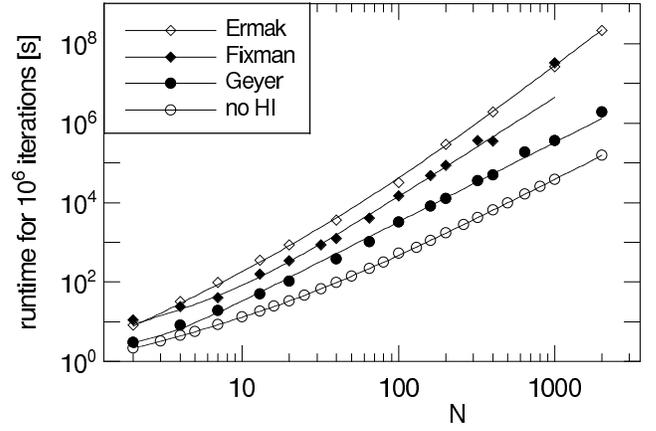}
	\end{center}
	\caption{Comparison of the runtime behavior of BD simulations with
	the different methods to include HI vs. the chain length $N$ of
	the bead-spring polymer. The datapoints give the times required to
	simulate one million timesteps, the lines are polynomial fits to
	the data as explained in the text.}
	\label{fig:Laufzeit}
\end{figure}

Without HI, one needs to evaluate at each timestep the forces from the
$N-1$ springs connecting the $N$ beads and also the $N(N-1)/2$
repulsive two-body interactions. For each of the $N$ beads a random
displacement has to be chosen and then the beads are moved according
to the external and the random forces. Consequently, the observed
runtime $T$ per timestep can be fitted with $T = t_0 + t_1\,N +
t_2\,N(N-1)$ as shown in figure \ref{fig:Laufzeit}. For one million
timesteps on a 2.8 GHz Pentium 4 CPU, we obtained $t_0 = 0.2$ seconds,
$t_1 = 0.95$ seconds, and $t_2 = 0.038$ seconds. To obtain the
runtimes, we ran each of the simulations for several minutes until the
runtime for one million timesteps could be calculated with sufficient
statistical accuracy.

With HI, also the hydrodynamic tensor has to be set up, for which
$N(N-1)/2$ distances between the beads have to be calculated. For our
truncated expansion, the $N$ normalization constants $C_i$
\eq{eq:Normalisation} are evaluated in $\mathcal{O}(N^2)$ time. For
the expansion coefficients $\beta_{ij}$, the average coupling
$\epsilon = \langle D_{ij} / D_{ii}\rangle$ is required, for which the
$(N-1)(N-2)/2$ off-diagonal entries of the diffusion tensor have to be
summed up. Finally, the effective hydrodynamically corrected forces of
equations \eq{eq:Feffective} and \eq{eq:Ansatz} are evaluated in
$\mathcal{O}(N^2)$ time, leading to an overall quadratic runtime that
can be fitted with $t_0 = 1$ seconds, $t_1 = 0.2$ seconds, and $t_2 =
0.3$ seconds for one million timesteps. For the very simple polymer
model used, the inclusion of HI thus slowed down the simulations by a
constant factor of about ten for large $N$. For more realistic ---and
thus more expensive---interactions between the individual beads, the
relative cost of considering HI will even decrease.

In the original formulation of Ermak and McCammon \cite{ERM78}, the
random displacements are determined via a Cholesky factorization of
the diffusion tensor. This led to the expected runtime behavior of
$\mathcal{O}(N^3)$ in our simulations, which was fitted with $T =
0.027 N^3 + 1.5 N^2 + 0.8 N$ seconds for one million timesteps (see
figure \ref{fig:Laufzeit}). The more efficient Chebyshev expansion led
to an effective runtime of $\mathcal{O}(N{^{2.5}})$, i.e., $T = 0.14
N^{2.5} + 4N$ at an accuracy of $\epsilon_f = 10^{-3}$. Reducing
$\epsilon_f$ to $10^{-2}$ or $10^{-1}$ led to a small speedup of a
factor of 2 or 3, respectively. We note that the Chebyshev expansion
needs a lot of computer memory for the repeated matrix-vector
multiplications, especially for large particle numbers. Consequently,
at $N=1000$ the Chebyshev approximation was slowed down due to memory
constraints of the computer we used and the runtime for $N=2000$ could
not be determined any more on this machine. For our approximation, the
memory requirements were determined by the diffusion tensor, while the
Cholesky factorization needed additional temporary storage of about the
same size as the diffusion tensor.

With the very simple interactions between the beads that we used here
in the simulations, the inclusion of hydrodynamics with our
approximation slowed down the simulation by a constant factor of about
ten. A simulation of a polymer of 1000 beads would consequently run
for about half of a day with our approximated HI for every hour it
takes without HI. The same simulation would take four to five days
with Fixman's Chebyshev approximation and more than a month with the
original algorithm of Ermak and McCammon.

The propagation was performed with the simple Euler propagator and a
rather conservative timestep. The simulations therefore could be
easily accelerated by using a more advanced propagation scheme like,
e.g., the semiimplicit scheme of Jendrejack \cite{JEN00} or the 
Trotter expansion of De Fabritiis et al. \cite{FAB06}.

As our approximation calculates the effective random forces from
individual two-body contributions, it is now also possible to
introduce distance dependent cut-offs to the hydrodynamic interactions
to reduce the number of terms that have to be summed up. Similar to
the cut-offs for direct interactions, this will speed up the
simulation. With the traditional methods of Ermak and McCammon or of
Fixman, a cut-off would lead to vanishing entries in the diffusion
tensor without reducing its size and, thus, the numerical effort to
factorize it. With these algorithms, cut-offs only introduce errors in
the dynamics without any gain. We did not consider cut-offs here, as
it is a project on its own to investigate the trade-off between the
perturbations of the hydrodynamic interactions vs. the achieved
runtime savings.

Finally, we note that with the two-body contributions used in our
TEA-HI, the hydrodynamic coupling can be summed up in parallel
to the inter-particle forces. Then, there is no need to build up and
keep the complete $3N \times 3N$ hydrodynamic tensor in computer
memory, only the actually required $3 \times3$ submatrices
$\mathbf{D}_{ij}$ of equations \eq{eq:RPYfern} and \eq{eq:RPYnah} have
to be determined temporarily. Then the required memory for
many-particle simulations increases only linearly with $N$.
Consequently, our truncated expansion is not only faster than the
originally proposed Cholesky factorization and the improved Chebyshev
expansion, but it also needs much less computer memory, which allows
for even larger systems to be handled efficiently.

\section{Conclusions}

We showed how the correlated random displacements can be approximated
efficiently in Brownian dynamics simulations with hydrodynamic
interactions. In our truncated expansion approximation, effective
hydrodynamically corrected random forces are determined with the aim
to reproduce the statistical moments of the thermal motion. Truncating
the expansion at the second order allows to calculate the random
forces from individual two-body contributions in $\mathcal{O}(N^2)$
time. For these only $3 \times 3$ submatrices of the $3N \times 3N$
hydrodynamic tensor are required temporarily. Our approximation
consequently has the same runtime and storage scaling as the
calculation of the direct interactions between the particles.

We then compared the simulation results with our approximation to the
results obtained with the exact method of Ermak and McCammon and with
the Chebyshev approximation of Fixman, the runtimes of which scaled as
$\mathcal{O}(N^3)$ and $\mathcal{O}(N^{2.5})$, respectively. 

Both the dynamics of a bead-spring dimer with variable separation of
the beads and of polymers of chain lengths of $N = 2 \ldots 200$
showed that our approximation captures about 95\% of the hydrodynamically
introduced correlation. For most applications on chemical or biological
systems, where the direct interactions have to be approximated anyway,
this appears a completely sufficient level of accuracy. Moreover, it 
allows for accounting of important hydrodynamic effects in systems 
where they were sofar mostly ignored for computational reasons.

As already mentioned above, we see the main application of our
truncated expansion hydrodynamics in Brownian dynamics simulations of
large biological systems, where one is interested in the dynamics of
many-particle association \cite{GOR04}, transport processes, or
protein folding where the complicated interactions between the
proteins or parts thereof have to be approximated rather crudely. For
these applications, it is surely better to include most of the effects
of hydrodynamics without slowing down the simulation too much than to
either have no HI at all or the correct HI at a prohibitive
computational cost. Other applications could be to describe the
behavior of non-spherical particles by assembling them from smaller
spheres \cite{DIC84, GAR81}. Here the quality of the model increases
with the number of spheres used to build the non-spherical particles.

\newpage

\end{document}